\documentclass[11pt]{article}

\usepackage{amscd,amsthm, amsmath, amsfonts,amssymb}
\voffset= - 1.2in
\hoffset= - 1.0in

\title{Determinant Formulas for Matrix Model Free Energy}
\author{{\bf D.Vasiliev}\thanks{E-mail: \ vasiliev@itep.ru}
\date{ } \\ {\small {\it ITEP, Moscow, Russia}~ and
{\it MIPT, Moscow, Russia}}
}

\begin{document}
\maketitle

\vspace{-5cm}
\begin{center}
\hfill ITEP/TH-44/05\\
\end{center}
\vspace{5cm}

\abstract{The paper contains a new non-perturbative representation for subleading contribution to
the free energy of multicut solution for hermitian matrix model. This representation is a generalisation of the formula,
proposed by Klemm, Marino and Theisen for two cut solution, which was obtained by comparing the cubic matrix model 
with the topological B-model on the local Calabi-Yau
geometry $\widehat {II}$ and was checked perturbatively. In this paper we give a direct proof of their formula and generalise it 
to the general multicut solution.}

\section{Introduction}

An interest to the multicut solutions to matrix models was inspired by the studies in
${\cal N}=1$
supersymmetric gauge
theories due to Cachazo, Intrilligator and Vafa \cite{Cachazo:2001jy} and Dijkgraaf, Vafa  
\cite{Dijkgraaf:2002fc,Dijkgraaf:2002vw,Dijkgraaf:2002dh}  who proposed to calculate the nonperturbative
superpotentials of ${\cal N}=1$ SUSY gauge theories in
four dimensions using matrix models technique. 
This ${\cal N}=1$ theories contains the multiplet of ${\cal N}=2$ SUSY gauge theories 
but with nontrivial tree superpotential.
The nonperturbative superpotential could be obtained from the partition
functions of the one-matrix model (1MM) in the leading order in
$1/N$, $N$ being the matrix size.
Higher genus corrections are identified with certain holomorphic couplings of gauge theory to gravity.

The authors of \cite{Klemm:2002pa} proposed a new anzatz for ${\cal F}_1$ in the two-cut
case (with absent double points) and made a perturbative check. Their formula in fact comes from the
correspondence between the so called topological B-model on the local Calabi-Yau
geometry $\widehat {II}$ and the cubic
matrix model. Here we give complete proof of this formula and generalize it to the multi-cut case.

We start with definition of the matrix integral and introduce all relevant constructions. For a complete review of the subject,
see \cite{Chekhov:2005kd} and references there in. 
Consider the hermitian 1-matrix model:
\begin{align}
\int_{N\times N}DX\, e^{-{\frac1\hbar}tr V(X)}=e^{\cal F},
\label{Xap2.1}
\end{align}
where $V(X)=\sum_{n\geq 1}^{}t_nX^n$, $\hbar = {\frac{t_0}N}$ is
a formal expansion parameter, the integration goes
over the $N\times N$ matrices, $DX\propto\prod_{ij}dX_{ij}$ 

The topological expansion of the Feynman diagrams series is then equivalent to
the expansion in even powers of $\hbar$ for
\begin{align}
{\cal F}\equiv {\cal F}(\hbar,t_0, t_1, t_2, \dots)
=\sum_{h=0}^{\infty}{\hbar}^{2h-2}{\cal F}_h,
\label{Xap2.2}
\end{align}
Customarily $t_0=\hbar N$ is the scaled
number of eigenvalues. We assume the potential $V(p)$ to be a polynomial
of the fixed degree $m+1$.

The averages, corresponding to the partition function (\ref{Xap2.1}) are
defined as usual:
\begin{align}
\bigl\langle f(X)\bigr\rangle=
\frac1Z\int_{N\times N}DX\,f(X)\,\exp\left(-{\frac1\hbar}tr V(X)\right)
\label{4.1}
\end{align}
and it is convenient to use their
generating functionals: the one-point resolvent
\begin{align}
W(\lambda)&=
\hbar
\sum_{k=0}^{\infty}
\frac{\langle tr X^{k}\rangle}{\lambda^{k+1}}.
\label{4.2}
\end{align}
as well as the $s$-point resolvents $(s\geq2)$
\begin{align}
W(\lambda_1,\dots,\lambda_s)&=\nonumber\\
\hbar^{2-s}
\sum_{k_1,\dots,k_s=1}^{\infty}
\frac{\langle tr X^{k_1}\cdots tr X^{k_s}\rangle_{\mathrm{conn}}}
{\lambda_1^{k_1+1}\cdots \lambda_s^{k_s+1}}&=\nonumber\\
\hbar^{2-s}
\left\langle tr\frac{1}{\lambda_1-X}\cdots
tr\frac{1}{\lambda_s-X}\right\rangle_{\mathrm{conn}}
\label{4.3}
\end{align}

The genus expansion of the resolvent has the form
\begin{align}
W(\lambda_1,\dots,\lambda_s)=\sum_{h=0}^{\infty}
\hbar^{2h}
W_{h}(\lambda_1,\dots,\lambda_s),\quad s\geq1,
\label{4.7}
\end{align}
It satisfies the loop equation \cite{Makeenko:1991tb,Migdal:1984gj}:

\begin{align}
\left[V'(x)W(x)\right]_-=
W(x)^2+
\hbar^2
W(x,x),
\label{4.8}
\end{align}
where $\left[...\right]_-$ is the projector on the negative powers. In genus zero, loop equations have the solution

\begin{align}
W_0(\lambda)&= \frac{1}{2}(V'(\lambda) - y)\\
y^2&=V'(\lambda)^2+4P_{m-1}(\lambda),
 \label{*loop2*}
\end{align}
where $P_{m-1}$ is an arbitrary polynomial of degree $m-1$.
If the curve (\ref{*loop2*}) has $n$ cuts, it can be represented in terms of branching points $\mu_\alpha$

\begin{align}
\label{ty}
y\equiv M(\lambda)\tilde y\equiv M(\lambda)\sqrt{\prod\nolimits_{\alpha=1}^{2n}(\lambda-\mu_\alpha)}.
\end{align}
In this article we concentrate on the case with $m=n$ (without double points, i.e. $M(\lambda)$ is a constant). 
Thus the full set of moduli is: $t_I\equiv\{S_i,~t_0,~t_k\}$, $i=\overline{1,n-1}$, $k=\overline{1,n}$, where 
occupancy numbers $S_i$ are defined as integrals over A-cycles on the curve $y$,

\begin{align}
\label{SWa}
S_i\equiv\frac1{4\pi i}\oint_{A_i}yd\lambda
\end{align}

To construct ${\cal F}_1$, we also define the polynomials $H_I(\lambda)$

\begin{align}
\label{dOI}
{\frac{d y}{d t_I}}=
{ \frac{H_{I}(\lambda)}{y(\lambda)}}
\end{align}
and matrix $\sigma_{i,j}$

\begin{align}
\label{Q}
\sigma_{j,i}\equiv\oint_{A_j}\frac{\lambda^{i-1}}{y
(\lambda)}d\lambda,\quad i,j=\overline{1,n-1}.
\end{align}
It can be shown \cite{Chekhov:2005kd} that for polynomials 
$ H_k(\lambda)\equiv \sum_{l=1}^{n-1} H_{l,k}\lambda^{l-1}$, \ $k=\overline{1,n-1}$ corresponded $S_k$, 

\begin{align}
\label{inverse}
\sum_{l=1}^{n-1} \sigma_{j,l} H_{l,k}=\delta_{j,k}\quad\hbox{for}\quad j,k=\overline{1,n-1}.
\end{align}

\section{Two-cut case}

According to paper \cite{Klemm:2002pa} the holomorphic part of the genus one B-model amplitude is, up to an additive constant,
\begin{align}\label{Theisen}
{\cal F}_1=\frac 1 2 \log{\left(\det\left(\frac{\partial \mu^-_i} {\partial S_j}\right)\Delta^{2/3}\frac 2 {\mu^+_2-\mu^+_1}\right)},
\end{align}
where $\mu^-_1=\frac12(\mu_1-\mu_2)$, $\mu^-_2=\frac12(\mu_3-\mu_4)$, $\mu^+_1=\frac12(\mu_1+\mu_2)$ and $\mu^+_2=\frac12(\mu_3+\mu_4)$.
On the other hand there is an answer for ${\cal F}_1$ obtained directly from solving the loop equations (\ref{4.8}) for matrix model 
\cite{Akemann:1996zr,Chekhov:2004vx}, or using conformal field theory technique \cite{Kostov:1999xi,Dijkgraaf:2002yn}
\begin{align}\label{Akemann}
{\cal F}_1=-\frac 1 {24} \log\left(\prod_{\alpha=1}^{2n}M(\mu_\alpha)\cdot\Delta^{4}\cdot
(\det_{i,j}\sigma_{j,i})^{12}\right),
\end{align}
which, in the two-cut case without double points reads as
\begin{align}\label{Akemann2}
{\cal F}_1=-\frac 1 {24} \log\left(\Delta^{4}\cdot
\sigma^{12}\right),
\end{align}
$\sigma$ (\ref{Q}) here is $1\times1$ matrix. 
To obtain (\ref{Theisen}) from (\ref{Akemann2}), one should prove the following formula
\begin{align}\label{Stat}
\det\left(\frac{\partial \mu^-_i} {\partial S_j}\right)\Delta\frac 2 {\mu_3+\mu_4-\mu_1-\mu_2}\sigma=1.
\end{align}
We can explicitly find the derivatives $\frac{\partial S_i} {\partial \mu^-_j}$ (instead of $\frac{\partial \mu^-_i} {\partial S_j}$), 
keeping times $t_k$ constant. To do so one should first write $\frac{\partial S_i} {\partial \mu_j}$ then make the 
change of variables from $\{\mu_1$, $\mu_2$, $\mu_3$, $\mu_4\}$ to
$\{t_1$, $t_2$, $\mu^-_1$, $\mu^-_2\}$. 
Then 
\begin{align}\label{Variables}
\frac{\partial S_i}{\partial \mu^-_j}=
\frac{\partial S_i}{\partial \mu_k}\frac{\partial \mu_k}{\partial\mu^-_j},~i,j=\overline{1,n},~k=\overline{1,2n}.
\end{align}
$\frac{\partial \mu_k}{\partial\mu^-_j}$ here are obtained by inverting the matrix $\left(\frac{\partial\mu^-_j}{\partial \mu_k},
\frac{\partial S_j}{\partial \mu_k}\right)$.
After this, it is easy to rewrite (\ref{Stat}) using the elliptic integrals:
\begin{align}\label{Elliptic_Stat}
\sqrt\frac{\mu_4-\mu_2} {\mu_3-\mu_1}\left(\frac{\mu_4 - \mu_1} {\mu_4-\mu_2} \Pi\left(-\frac{\mu_2-\mu_1}{\mu_4-\mu_2},\kappa\right)\right.
+\nonumber
\\\left.\frac{\mu_3-\mu_2} {\mu_4-\mu_2}  \Pi\left(\frac{\mu_4-\mu_3}{\mu_4-\mu_2},\kappa\right)+  K(\kappa)\right)
&={\frac \pi 2}.
\end{align}
where
$\kappa=\sqrt{\frac{(\mu_2-\mu_1)(\mu_4-\mu_3)}{(\mu_4-\mu_2)(\mu_3-\mu_1)}}$, $\Pi(\nu,\kappa)$ and $K(\kappa)$
are complete elliptic integrals of the third and first kinds respectively.
To prove this statement, one can rewrite the elliptic integrals of the third kind via the 
complete and incomplete elliptic integrals of the first and the second kinds 
(these formulas can be found in \cite{Bateman} (formulas 22, 24 from chapter 13.8); 
note, however, that in \cite{Bateman} there is a misprint in these formulas)
\begin{align}
\label{Beitman22}
k'^2\frac{\sin\theta \cos\theta}{\sqrt{1-k'^2{\sin\theta}^2}}[\Pi(1-k'^2{\sin\theta}^2,\kappa)-K(\kappa)]&=\nonumber\\
\frac\pi2-(E(\kappa)-K(\kappa))F(\sin\theta,k')-K(\kappa)E(\sin\theta,k')\\
\label{Beitman24}
\frac{\sqrt{1-k'^2{\sin\theta}^2}}{\sin\theta \cos\theta}[\Pi(-k'^2{\tan\theta}^2,\kappa)-K(\kappa){\cos\theta}^2]&=\nonumber\\
(E(\kappa)-K(\kappa))F(\sin\theta,k')-K(\kappa)E(\sin\theta,k')
\end{align}
where $k'=\sqrt{1-k^2}$, $\theta\in[0,\pi/2]$. 
In this case one should put $\sin^2\theta=\frac{\mu_3-\mu_1}{\mu_4-\mu_1}.$

The same computation can be done for any other partition of $\mu_i$ into the two sets $\mu^\pm_{1,2}$ (without changing $\sigma$),
say, for
$\mu^-_1=\frac12(\mu_1-\mu_3)$, $\mu^-_2=\frac12(\mu_2-\mu_4)$, $\mu^+_1=\frac12(\mu_1+\mu_3)$ and $\mu^+_2=\frac12(\mu_2+\mu_4)$. 
It leads to the same result (\ref{Theisen}), however, the perturbative calculation in this case is irrelevant. 

\section{Generalization for n-cut solution}

A natural generalisation of (\ref{Theisen}) 
is
\begin{align}
{\cal F}_1=\frac12
\log\left(\det\left\|\frac{\partial \{\mu_j^-\}}{\partial\{S_i,S_n\}}\right\|
\Delta^{2/3}\Delta^{-1}(\mu_j^+)\right),
\label{F1brane}
\end{align}
where we divided all the branching points into two ordered sets
$\{\mu^{(1)}_j\}_{j=1}^n$ and $\{\mu^{(2)}_j\}_{j=1}^n$ and
performed a linear orthogonal transformation of
$\mu_j^{(1,2)}$ to the quantities
$\{\mu_j^+\}$ and $\{\mu_j^-\}$, $j=\overline{1,n}$,
\begin{align}
\mu_j^{\pm}=\mu^{(1)}_{j}\pm\mu^{(2)}_{j}.
\label{**1*}
\end{align}

To prove formula (\ref{F1brane}), one should calculate the derivative of the branching points $\mu_j$ with respect 
to the moduli $\{t_K\}\equiv\{S_1..S_{n-1},t_0..t_n\}$ \cite{Chekhov:2005kd}:

\begin{align}
\label{diff2}
\frac{\partial \mu_\alpha}{\partial t_K}=\frac{ H_{K}(\mu_\alpha)}
{M(\mu_\alpha)\prod_{\beta\ne\alpha}
(\mu_\alpha-\mu_\beta)}.
\end{align}

The polynomials $H_{I}(\lambda)$ corresponding to the variables $t_k,~k\geq 1$
always have the coefficient $k$ at the highest term 
$\lambda^{n-1+k}$ and the polynomial corresponding to $t_0$ 
starts with $\lambda^{n-1}$. Therefore, one can find the determinant:

\begin{align}
\det\left\|\frac{\partial \{\mu_{\alpha_j}\}}{\partial \{S_i,S_n,t_k\}}\right\|=
\frac{\Delta(\mu_{\alpha_j})\cdot \left(\det\limits_{l,m}\sigma_{l,m}\right)^{-1}}
{\prod\limits_{i=1}^{2n}M(\mu_{\alpha_i})\prod\limits_{j=1}^{2n}
\left(\prod\limits_{\beta\ne\alpha_j}^{2n}
(\mu_{\alpha_j}-\mu_\beta)\right)}
\label{**2*}
\end{align} 

Indeed, consider the left hand side of (\ref{**2*}). 
\begin{align}
\det\left\|\frac{\partial \{\mu_{\alpha_j}\}}{\partial \{S_i,S_n,t_k\}}\right\|=
\frac{\left(\det\limits_{K,j}H_K(\mu_{\alpha_j})\right)}
{\prod\limits_{i=1}^{2n}M(\mu_{\alpha_i})\prod\limits_{j=1}^{2n}
\left(\prod\limits_{\beta\ne\alpha_j}^{2n}
(\mu_{\alpha_j}-\mu_\beta)\right)}
\label{**2*left}
\end{align} 
The change of variables $\{S_1,..,S_{n}\}\rightarrow \{S_1,...,S_{n-1},t_0\}$ does not 
change the determinant. 
To obtain the Vandermonde determinant in the 
right hand side of (\ref{**2*}), there should be, instead of the polynomials
$H_K$, polynomials of degree
$2n-i+1$ where $i$ is the line number, with unit leading coefficients. 
To this end, one should multiply the matrix $H_{K}(\mu_{\alpha_{j}})$ with the block diagonal matrix
\begin{align}\tilde\sigma=\begin{pmatrix}
1&0\\
0&\sigma
\end{pmatrix}.\label{tildesigma}\end{align}
This gives the factor $(\det{\tilde\sigma})^{-1}=(\det\limits_{l,m}\sigma_{l,m})^{-1}$. 
Lines from $1$ to $n+1$ contribute to $n!$ which could be omitted from the free energy.
The Vandermonde determinant $\Delta(\mu_{\alpha_j})$ then combines with the rational
factors in the denominator to produce
$(-1)^{\sum_{j=1}^n\alpha_j}\Delta(\overline{\mu_{\alpha_j}})/\Delta(\mu)$, where
$\Delta(\overline{\mu_{\alpha_j}})$ is the Vandermonde determinant
for the supplementary set of $n$ branching points not entering
the set $\{\mu_{\alpha_j}\}_{j=1}^{n}$ whereas $\Delta(\mu)$
is the total Vandermonde determinant. Now we should put $M(\mu_\alpha)$ constant independent of $\alpha$. 
Expanding the determinant in (\ref{F1brane}) by each line and neglecting the 
additive constant $\frac12\log{2^n}$,  one obtain (\ref{Akemann}).

Introducing the quantities 
\begin{align}\label{phi}
\phi_I^{\alpha}\equiv
\frac{ H_I(\mu_{\alpha})
}
{M^{1/3}(\mu_\alpha)
\prod_{\beta\ne\alpha}(\mu_{\alpha}-\mu_{\beta})^{2/3}},
\end{align}
one can rewrite (\ref{Akemann}) in a more simple form:
\begin{align}\label{F-phi}
{\cal F}_1=\frac12\log\left(\det_{I,\alpha}\phi_I^{\alpha}\right)\end{align}

\section{Perturbative Formula}

We have also performed the perturbative check of (\ref{F1brane}) for the 3-cut case.
It is easier to make the expansion not in the moduli $S_i$ but in the difference of the branching points $\mu^-_j$.
In order to calculate $\det\left\|\frac{\partial\{S_i,S_n\}}{\partial \{\mu_j^-\}}\right\|$, one should rewrite $S_i$ and $\sigma_{i,j}$
in terms of $\mu^+_i$, $\mu^-_j$ and expand them in $\mu^-_i$
\begin{align}
\label{S_pert}
S_l& = \frac12 res_{\lambda=\mu_l^+}\prod^n_{i=1}\sum^\infty_{k=0}\frac{(\mu^-_i)^{2k}c_k}{(\lambda-\mu^+_i)^{2k-1}}\\
\label{Sigm_pert}
\sigma_{l,j}& = \frac12 res_{\lambda=\mu_l^+}\lambda^{j-1}\prod^n_{i=1}\sum^\infty_{k=0}
\frac{(\mu^-_i)^{2k}\tilde c_k}{(\lambda-\mu^+_i)^{2k+1}}
\end{align} 
$c_k$ and $\tilde c_k$ are the Taylor coefficients for $\sqrt{1-x}$ and $\frac1{\sqrt{1-x}}$ respectively.
It should be mentioned that derivatives $\frac{\partial\{S_i\}}{\partial \{\mu_j^-\}}$ are taken at $t_k$ constant, while in
(\ref{S_pert})  
$S_k$ are functions of $\mu^+,\mu^-$. 
This problem is solved by calculating the transition matrix from 
$\{\mu^-_k,t_k\}$ to $\{\mu^-_k,\mu^+_k\}$ and inverting it. We have done this calculation up to 
$(\mu^-)^3$ and found it in perfect agreement with (\ref{Akemann}) (up to an additive constant mentioned).

\section*{Acknowledgments}

Our work is partly supported by Federal Program of the Russian Ministry of
Industry, Science and Technology No 40.052.1.1.1112 and by the grant
RFBR 04-02-16880


\end{document}